\documentstyle [12pt,epsf]{article}
 \newcommand{\insertplot}[5]{\begin{figure}
 \hfill\hbox to 0.05in{\vbox to #5in{\vfill
 \inputplot{#1}{#4}{#5}}\hfill}
 \hfill\vspace{-.1in}
 \caption{#2}\label{#3}
 \end{figure}}
 \newcommand{\inputplot}[3]{
 \special{ps: plotfile #1}
}

\begin{document}
 
\title{Dyons with axial symmetry}
\vspace{1.5truecm}
\author{{\bf Betti Hartmann$^1$,
               Burkhard Kleihaus$^2$
               and Jutta Kunz$^1$ }\\
  $^1$Fachbereich Physik, Universit\"at Oldenburg, Postfach 2503\\
      D-26111 Oldenburg, Germany\\
  $^2$ Dublin University College\\
      Dublin, Ireland}
 
\date{\today}

\maketitle

\begin{abstract}
We construct axially symmetric dyons in SU(2) Yang-Mills-Higgs theory.
In the Prasad-Sommerfield limit,
they are obtained via scaling relations from 
axially symmetric multimonopole solutions.
For finite Higgs self-coupling they are constructed numerically.
\end{abstract}

\section{Introduction}

SU(2) Yang-Mills-Higgs (YMH) theory 
possesses magnetic monopole \cite{mono1,ps,mono3},
multimonopole \cite{multi1,multi2,kkt}
and monopole-antimonopole solutions \cite{anti1,anti2,kk}.
The magnetic charge $m$ of these solutions is proportional
to their topological charge $n$.
In the Prasad-Sommerfield limit of
vanishing Higgs self-coupling,
the monopole solution and axially symmetric multimonopole solutions
are known analytically \cite{ps,multi2},
whereas multimonopole solutions without
rotational symmetry \cite{norot}
and monopole-antimonopole solutions \cite{anti2,kk}
are only known numerically.

SU(2) Yang-Mills-Higgs (YMH) theory 
also possesses solutions carrying both magnetic and electric charge
\cite{jul,ps}.
In the Prasad-Sommerfield limit, 
such dyon solutions with unit topological charge
exist for arbitrarily large values of the electric charge \cite{ps},
whereas for finite Higgs self-coupling
an upper bound for
the electric charge of these dyon solutions arises \cite{bkt}.

Here we construct axially symmetric dyon solutions
carrying topological charge $n>1$.
In section 2 we present the YMH Lagrangian, 
the equations of motion and the electromagnetic charges.
In section 3 we consider the Prasad-Sommerfield limit,
and show how dyon solutions are obtained from multimonopole solutions.
We present the axially symmetric ansatz and the boundary conditions
employed in the numerical construction of the
dyonic multimonopole solutions in section 4,
and discuss these solutions in section 5, in particular
for finite Higgs self-coupling.

\section{SU(2) YMH Theory}

We consider the SU(2) YMH Lagrangian
\begin{equation}
L = -\frac{1}{4} F_{\mu\nu}^a F^{\mu\nu,a} 
    -\frac{1}{2} D_\mu \Phi^a D^\mu \Phi^a 
      - \frac{\lambda}{4} (\Phi^a \Phi^a - \eta^2)^2 \ ,
\label{lag}
\end{equation}
with
\begin{equation}
F^a_{\mu\nu} = \partial_\mu A^a_\nu - \partial_\nu A^a_\mu
 + g \varepsilon^{abc} A^b_\mu A^c_\nu
\ , 
\label{fmunu}
\end{equation}
\begin{equation}
D_\mu \Phi^a =
  \partial_\mu \Phi^a +g \varepsilon^{abc} A_\mu^b \Phi^c
\ , 
\label{dmuphi}
\end{equation}
gauge coupling constant $g$,
Higgs self-coupling constant $\lambda$,
Higgs vacuum expectation value $\eta$,
and equations of motion
\begin{eqnarray}
D_\mu F^{\mu\nu,a} & = &- g \varepsilon^{abc} (D^\nu \Phi^b) \Phi^c \ ,
\nonumber \\
D_\mu D^\mu \Phi^a & = & - \lambda \left( \Phi^b \Phi^b 
  - \eta^2 \right) \Phi^a \ .
\label{feq1}
\end{eqnarray}

We are looking for static finite energy solutions of eqs.~(\ref{feq1}),
carrying both magnetic and electric charge.
The magnetic field $B_i$ and the electric field $E_i$
are defined via the electromagnetic `t Hooft field strength tensor
\cite{mono1}
\begin{equation}
{\cal F}_{\mu\nu} = \hat{\Phi}^a F_{\mu\nu}^a 
-\frac{1}{g} \varepsilon_{abc} \hat{\Phi}^a
 D_\mu \hat{\Phi}^b D_\nu \hat{\Phi}^c \ ,
\end{equation}
where in particular
\begin{equation}
E_i = {\cal F}_{i0} = \partial_i(\hat{\Phi}^a A_0^a) \ .
\label{ef} \end{equation}
For the dyon solutions they yield the magnetic charge $m$
\cite{raja},
\begin{equation}
 m  =  \frac{1}{4\pi}\int{\partial_i B_i d^3r}  = \frac{n}{g} \ ,
\label{cm} \end{equation}
where $n$ is the topological charge of the solutions, and
the electric charge $q$ 
\begin{equation}
 q  =  \int{\partial_i E_i d^3r} \  
= \int_{S_\infty^2} \partial_i(\hat{\Phi}^a A_0^a)dS_i
\label{ce} \label{charges}
 \end{equation}
where $S^2_\infty$ denotes the 2-dimensional sphere at infinity.

The energy density of the dyon solutions is given by the
$tt$-component of the energy momentum tensor. Integration
over all space yields their energy
\begin{equation}
E = \int{\left\{\frac{1}{4} F_{ij}^a F_{ij}^a+\frac{1}{2} F_{i0}^a F_{i0}^a 
+\frac{1}{2} D_i \Phi^a D_i \Phi^a+\frac{1}{2} D_0 \Phi^a D_0 \Phi^a 
+\frac{\lambda}{4} (\Phi^a \Phi^a - \eta^2)^2 
\right\} d^3r} \ .
\label{engd1}
\end{equation}  

\boldmath
\section{Prasad-Sommerfield limit $\lambda=0$}
\unboldmath

Let us first consider the Prasad-Sommerfield limit $\lambda=0$.
In this limit axially symmetric multimonopole solutions are
known analytically \cite{multi2} and numerically \cite{kkt}.

We now show, that in the static limit
dyon solutions are obtained directly
from monopole solutions. 
Assuming that the time component of the 
gauge field is parallel to the Higgs field in isospace \cite{wein},
\begin{equation}
A_0^a = \alpha \Phi^a \ ,
\label{a0}
\end{equation}
where $\alpha$ is a constant,
leads to
$D_0\Phi^a=0$ and 
$F_{i0}^a =\alpha (\partial_i \Phi^a +g \varepsilon_{abc} A_i^b \Phi^c)
= \alpha D_i \Phi^a$.
Consequently, the field equations (\ref{feq1}) reduce to
\begin{eqnarray}
D_i F^{ij,a} & = &- g \varepsilon_{abc} (1-\alpha^2) D^j \Phi^b \Phi^c \ ,
\nonumber \\
D_i D^i \Phi^a & = & 0 \ ,
\label{feq}
\end{eqnarray}
and the field equation for the time component of the gauge field coincides 
with the field equation for the Higgs field.
Substituting $\tilde{\Phi}^a = \sqrt{1-\alpha^2}\Phi^a$ into eqs.~(\ref{feq})
leads to the monopole equations.

Thus, to any static solution $(A_i^a,\tilde{\Phi}^a)$ of the monopole 
equations in the Prasad-Sommerfield limit with 
$|| \tilde{\Phi}^a || \stackrel{r \rightarrow \infty}{\longrightarrow} 
\tilde{\eta}$ there corresponds a family of dyon solutions 
\begin{equation} \left(A_i^a\ , 
\ \   A_0^a=Q\tilde{\Phi}^a \ , 
\ \ \Phi^a =\sqrt{1+Q^2}\tilde{\Phi}^a \right) \ ,
\label{dyons}
\end{equation}
with 
${\displaystyle || \Phi^a || \stackrel{r \rightarrow \infty}{\longrightarrow} 
\tilde{\eta}\sqrt{1+Q^2}}$, where we have introduced 
\begin{equation}
Q=\alpha/\sqrt{1-\alpha^2} \ . \label{Qalpha}
\end{equation}

Let us now turn to the energy of the dyon solutions.
If $(A_i^a,\tilde{\Phi}^a)$
is a solution of the monopole equations, it extremizes the energy
functional 
\begin{equation}
E_{{\rm MP}} = \int{\left\{\frac{1}{4} F_{ij}^a F_{ij}^a
+\frac{1}{2} D_i \tilde{\Phi}^a D_i \tilde{\Phi}^a
\right\} d^3r } \ ,
\label{engMP}
\end{equation}
Since the scaling argument yields
\begin{equation}
\int{\left\{\frac{1}{2} D_i \tilde{\Phi}^a D_i \tilde{\Phi}^a\right\} d^3r} = 
\int{\left\{\frac{1}{4} F_{ij}^a F_{ij}^a\right\} d^3r} =
 \frac{1}{2} E_{{\rm MP}}
 \ ,
\end{equation}
we obtain for the energy of the dyon solutions 
(eq.~(\ref{dyons}) )
\begin{eqnarray}
E(Q) & = & \int{\left\{\frac{1}{4} F_{ij}^a F_{ij}^a
+\frac{1}{2} Q^2 D_i \tilde{\Phi}^a D_i \tilde{\Phi}^a
+\frac{1}{2} (1+Q^2)   D_i \tilde{\Phi}^a D_i \tilde{\Phi}^a
\right\} d^3r} 
\nonumber\\
 & = & \int{\left\{\frac{1}{4} F_{ij}^a F_{ij}^a
+(1+2Q^2)\frac{1}{2} D_i \tilde{\Phi}^a D_i \tilde{\Phi}^a
\right\} d^3r} 
\nonumber\\
 & = & 2(1+Q^2)\int{\left\{\frac{1}{4} F_{ij}^a F_{ij}^a\right\} d^3r} 
\nonumber\\
 & = & (1+Q^2)E_{{\rm MP}} \ .
\label{eemp}\end{eqnarray}

Note, that the above construction of 
electrically charged solutions is fairly general.
Applying it e.g.~to monopole-antimonopole solutions \cite{kk},
the resulting solutions possess 
magnetic charges of opposite sign, but electric charges of equal sign.
The reason is, that the magnetic charges are related 
to topological defects at the locations of the zeros of the 
Higgs field modulus, whereas the electric charge is related to 
the power law behaviour of the asymptotic Higgs field.

BPS monopole solutions satisfy
the (anti-)selfdual equations
\begin{equation}
F_{ij}^a = \pm\varepsilon_{ijk} D_k\tilde{\Phi}^a \ .
\label{BPS}
\end{equation}
Their energy saturates the lower bound 
$E_{{\rm MP}}= (4 \pi |n|/{g}) \tilde{\eta}$, where
$n$ denotes the topological charge.
According to eq.~(\ref{eemp}), we then find
for the energy of the dyon solutions
$ E(Q)=(4 \pi |n|/{g}) \tilde{\eta}(1+Q^2)$.
Choosing 
$\tilde{\eta}= \eta/{\sqrt{1+Q^2}}$, such that the Higgs field 
approaches asymptotically the value $\eta$, we obtain
\begin{equation}
 E(Q)=\frac{4 \pi |n|}{g} \eta\sqrt{1+Q^2} \ . \label{eQ1}
\end{equation} 
Moreover, the energy density 
$\epsilon(Q)$ of the dyon solutions is proportional to the energy density 
$\epsilon_{{\rm MP}}$
of the BPS solutions,
\begin{equation}
\epsilon(Q)=(1+Q^2)\epsilon_{{\rm MP}} \ .
\end{equation}

Let us finally express the energy of the dyon solutions in 
terms of their magnetic and electric charges.
With the electric field (\ref{ef})
\begin{equation}
E_i = \hat{\Phi}^a F_{i0}^a
= Q \hat{\tilde{\Phi}}^a \partial_i \tilde{\Phi}^a 
= Q \partial_i ||\tilde{\Phi}^a || \ 
\end{equation}
and the asymptotic form 
of the modulus of the Higgs field of BPS multimonopoles 
\cite{multi2}
\begin{equation}
||\tilde{\Phi}^a ||= \tilde{\eta}
\left( 1- \frac{|n|}{\tilde{\eta} g r}\right) + O(r^{-2})
\  
\end{equation}
the expression for the electric charge, eq.~(\ref{ce}), yields
\begin{equation}
q = Q \int_{S^2_\infty}
{\partial_i  ||\tilde{\Phi}^a ||}dS_i  
 =\frac{4 \pi |n| }{g} Q \ , \label{cee}
\end{equation}
Consequently, the energy of the dyon solutions is given by
\begin{equation}
E(m,q) = 4 \pi \eta \sqrt{m^2 +\left(\frac{q}{4\pi}\right)^2} \ .
\label{eQ2} \end{equation} 
In this form
the BPS expression for the energy reflects the electromagnetic
duality (see e.g.~\cite{ketov}).

\section{Axially Symmetric YMH Ansatz}

To construct dyon solutions numerically,
we employ the static axially symmetric ansatz
used in the numerical construction of multimonopoles \cite{multi1,kkt},
supplemented by a non-vanishing time-component of the gauge field.
In spherical coordinates with
$A_\mu^a dx^\mu = A_t^a dt + A_r^a dr + A_\theta^a d\theta + A_\varphi^a
 d\varphi$,
the ansatz reads
\begin{equation}
 A^a_t = \eta \left[ H_{5}(r,\theta)u_r^a + H_{6}(r,\theta) u_\theta^a 
 \right] \ ,
\end{equation}
\begin{equation}
 A^a_r= \frac{1}{r} H_{1}(r,\theta)u^a_\varphi \ , \ \ \
 A^a_\theta =  (1-H_{2}(r,\theta))u^a_\varphi \ ,
\end{equation}
\begin{equation}
 A^a_\varphi = -n \sin\theta \left[H_{3}(r,\theta)u^a_r +
(1-H_{4}(r,\theta))u^a_\theta \right] \ ,
\end{equation} 
\begin{equation}
\Phi^a = \eta \left[ \phi_{1}(r,\theta)u_r^a +
\phi_{2}(r,\theta)u^a_\theta \right] \ ,
\end{equation}
with unit vectors
\begin{equation}
\vec u_{r}= (\sin\theta \cos n\varphi,\sin\theta \sin n\varphi,
 \cos\theta) \ ,
\end{equation}
\begin{equation}
\vec u_{\theta}= (\cos\theta \cos n\varphi,\cos\theta \sin n\varphi,
 -\sin\theta) \ ,
\end{equation}
\begin{equation}
\vec u_{\varphi}= (-\sin n\varphi, \cos n\varphi, 0 ) \ .
\end{equation}
The winding number $n$ corresponds to the topological charge of the
solutions \cite{multi1,kkt}.
For $n=1$ the ansatz reproduces the spherically symmetric
dyons \cite{jul,bkt}.

Regularity, finite energy and symmetry requirements 
lead to the boundary conditions \cite{multi1,kkt}.
At the origin and at infinity they read
\begin{equation}
H_{i}(0,\theta)=0 \ , \ i=1,3,5,6 \ , \ \ \
H_{i}(0,\theta)=1 \ , \ i=2,4 \ , 
\end{equation}
\begin{equation}
\phi_{i}(0,\theta)=0 \ , \ i=1,2 \ ,
\end{equation}
\begin{equation}
H_{i}(\infty,\theta)=0 \ , \ i=1,2,3,4,6 \ , \ \ \
H_{5}(\infty,\theta)=\alpha \ ,
\label{jinf} \end{equation}
\begin{equation}
\phi_{1}(\infty,\theta)=1 \ , \ \ \
\phi_{2}(\infty,\theta)=0 \ ,
\end{equation}
and on the $z$- and $\rho$- axis (with $z=r \cos\theta$ and $\rho=r
\sin\theta$) they are
\begin{equation}
H_{i}(r,0)=0 \ , \ i=1,3,6 \ , \ \ \
\partial_{\theta}H_{i}(r,0)=0 \ , \ i=2,4,5 \ ,
\end{equation}
\begin{equation}
\partial_{\theta}\phi_{1}(r,0)=0 \ , \ \ \ \phi_{2}(r,0)=0 \ ,
\end{equation}
\begin{equation}
H_{i}(r,\pi/2)=0 \ , \ i=1,3,6 \ , \ \ \
\partial_{\theta}H_{i}(r,\pi/2)=0 \ , \ i=2,4,5 \ ,
\end{equation}
\begin{equation}
\partial_{\theta}\phi_{1}(r,\pi/2)=0 \ , \ \ \
\phi_{2}(r,\pi/2)=0 \ .
\end{equation}
The constant $\alpha$ in eq.~(\ref{jinf}) is restricted to 
$\alpha \le 1$. For $\alpha>1$
some gauge field functions become oscillating instead of
asymptotically decaying, completely analogous to the $n=1$ case
\cite{bkt}.

\section{Dyon Solutions}

We now present the axially symmetric dyon solutions, 
obtained numerically for vanishing and finite Higgs self-coupling.

\subsection{Vanishing Higgs self-coupling}

In the Prasad-Sommerfield limit
the time component of the gauge field and the Higgs field are proportional,
eq.~(\ref{a0}), in particular,
$H_{5}/\phi_{1}=\alpha$ and $H_{6}/\phi_{2}=\alpha$.
The proportionality constant $\alpha$, eq.~(\ref{Qalpha}),
then determines the electric charge, eq.~(\ref{cee}),
and the energy, eq.~(\ref{eQ1}) resp.~eq.~(\ref{eQ2}).

The numerically obtained axially symmetric solutions with 
topological charge $n=1$, 2 and 3
satisfy these relations very well.
Both the energy per topological charge 
$E_{n}/n$
and the electric charge
per topological charge 
$q_{n}/n$
are independent of the topological charge $n$
and depend only on the constant $\alpha$.
Since the electric charge diverges for $\alpha \rightarrow 1$,
the dyon solutions
exist for arbitrarily large values of the electric charge.

\subsection{Finite Higgs self-coupling}

For finite Higgs self-coupling, the simple scaling relations no longer hold.
Although for small Higgs self-coupling the time component of the gauge field
and the Higgs field are still almost proportional.

The magnetic charge is given in terms of the topological charge
also for finite Higgs self-coupling.
However, the electric charge is no more obtained from the asymptotic
behaviour of the
Higgs field $\Phi^a = ||\Phi^b|| \hat{\Phi}^a_\infty$ alone,
which now decays exponentially
\begin{equation}
||\Phi^a|| = \eta(1 -e^{-r\beta}) \ ,  \ \ \ \beta={\rm const} \ .
\label{decay} \end{equation}
Instead the asymptotic behaviour of the time component 
of the gauge field now primarily determines the electric charge,
eq.~(\ref{ce}),
\begin{equation}
q  = \lim_{r \rightarrow \infty}
4\pi \hat{\Phi}^a_\infty r^2 \partial_r A_t^a
= \lim_{r\rightarrow \infty}
4\pi \eta r^2 \partial_r H_5
\ . \end{equation}

In Fig.~1 we show the energy per topological charge 
$E_{n}/n$
in units of $4 \pi \eta /g$
for Higgs self-coupling $\lambda = 0.5$
and for topological number $n=1$, 2 and 3
as a function of the electric charge
per topological charge 
$q_{n}/n$
in units of $4 \pi /g$.
For dyons, as for magnetic monopoles \cite{mono3}, 
there exists only a repulsive phase for $\lambda \neq 0$,
because the attractive Higgs field becomes massive and thus 
exponentially decaying.
Therefore - unlike the BPS case - it can no longer 
cancel the long-range repulsive force of the gauge fields.

For finite $\lambda$
the energy per topological charge 
$E_{n}/n$
increases with increasing $n$.
The energy per topological charge ${E}_{n}/n$ of the
axialsymmetric solutions is higher than ${E}_{n}/n$
of the spherical symmetric solution. 
The axially symmetric $n=3$ solutions have higher
${E}_{n}/n$
than the corresponding $n=2$ solutions.
For $n \ge 3$ in addition
solutions with only discrete symmetries exist \cite{norot}.
We expect that for finite $\lambda$
these solutions are energetically more
favorable than the axially symmetric solutions
constructed here.

In contrast to the Prasad-Sommerfield limit, 
where dyon solutions exist for arbitrary electric charge,
there exists an ($n$-dependent) upper bound 
for the electric charge of dyon solutions
when the Higgs self-coupling is finite.
This is illustrated in Fig.~2, where we show
the electric charge
per topological charge $q_{n}/n$
as a function of the parameter $\alpha$
for $\lambda=0.5$ and $n=1$, 2 and 3.
For $\alpha \rightarrow 1$ the upper bound of the 
electric charge is reached.
The localized dyon solutions then
cease to exist, because some gauge field functions become oscillatory
instead of asymptotically exponentially decaying.
The upper bounds in Fig.~2, where $\alpha=1$, represent therefore
the endpoints of the curves in Fig.~1.

The upper bound of the electric charge of the dyon solutions
decreases with increasing Higgs self-coupling $\lambda$.
This is seen in Fig.~3 where we show
the electric charge
per topological charge $q_{n}/n$
as a function of the parameter $\alpha$
for $\lambda=0$, 0.5 and 1
for dyons with $n=2$.
Note, that the curve for $\lambda=0$ is independent of $n$.

The dyons with $n>1$ possess toroidal shape,
as is illustrated in Fig.~4.
In this three-dimensional plot the energy density 
of the dyon solution with topological charge $n=2$ and electric charge
per topological charge $q_{2}/2=0.6$ 
is shown 
for Higgs self-coupling $\lambda=0.5$
as a function of the
compactified coordinates $\rho = \bar x \sin \theta$ and
$z = \bar x \cos \theta$, where $\bar x = x/(1+x)$ is 
the coordinate used in the numerical calculations.
The effect of the presence of the electric charge is seen in Fig.~5,
where the energy density is shown as a function of
the compactified coordinate $\bar x$ for several values of
the angle $\theta$, both for the above dyon solution and for the
corresponding multimonopole solution.

\section{Conclusions}

We have constructed axially symmetric solutions of SU(2) YMH theory
carrying both magnetic and electric charge.
In the Prasad-Sommerfield limit solutions with electric
charge are obtained from purely magnetically charged solutions
by simple scaling relations.
The energy expression eq.~(\ref{eQ2})
for these BPS solutions reflects the electromagnetic
duality.

For finite Higgs self-coupling the energies don't satisfy
a duality relation analogous to eq.~(\ref{eQ2}).
In particular there exist $n$- and $\lambda$-dependent upper bounds
for the electric charge of dyon solutions,
where localized solutions cease to exist.

The toroidal shape of the energy density of the monopole solutions
is retained for the dyon solutions.
With increasing electric charge the maximum of the energy density decreases,
instead the energy density reaches further out.

As for the multimonopoles, dyonic solutions can be constructed for
the monopole-antimonopole solutions. 
These solutions then possess magnetic charges of opposite sign, 
but electric charges of equal sign.
In the Prasad-Sommerfield limit 
the dyonic monopole-antimonopole solutions can be obtained
by means of the same scaling relations as the dyonic
multimonopole solutions.
The numerical construction of dyonic monopole-antimonopole solutions
is presently under consideration.

{\bf Acknowledgement}

We gratefully acknowledge discussions with Y. Brihaye and T. Tchrakian.


\newpage

\begin{figure}
\centering
\epsfysize=11cm
\mbox{\epsffile{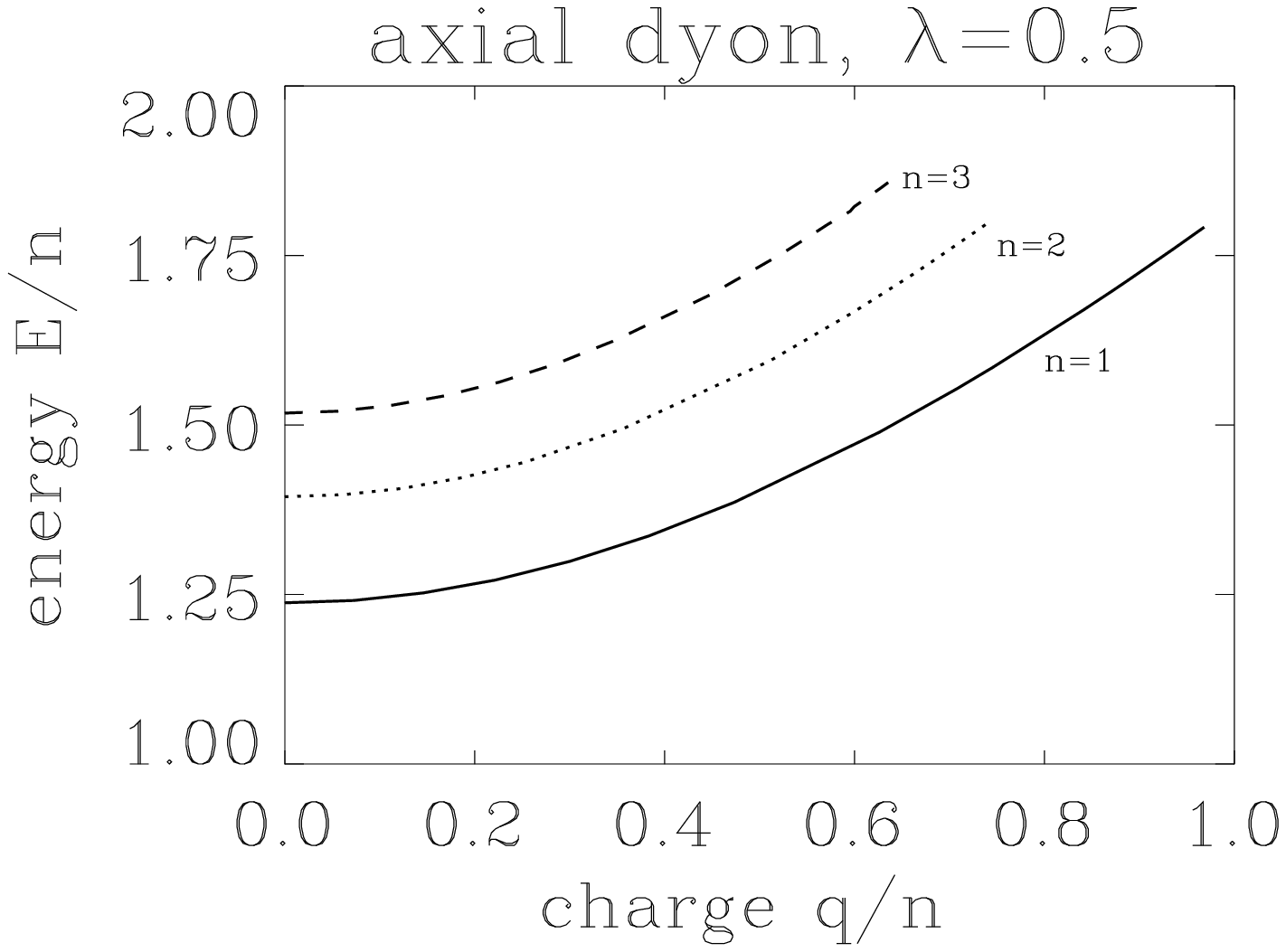}}
\caption{
The energy per topological charge $E_{n}/n$
in units of $4 \pi \eta /g$
is shown
as a function of the electric charge
per topological charge $q_{n}/n$
in units of $4 \pi /g$
for Higgs self-coupling $\lambda = 0.5$
and for topological number $n=1$, 2 and 3.
}
\end{figure}

\newpage

\begin{figure}
\centering
\epsfysize=11cm
\mbox{\epsffile{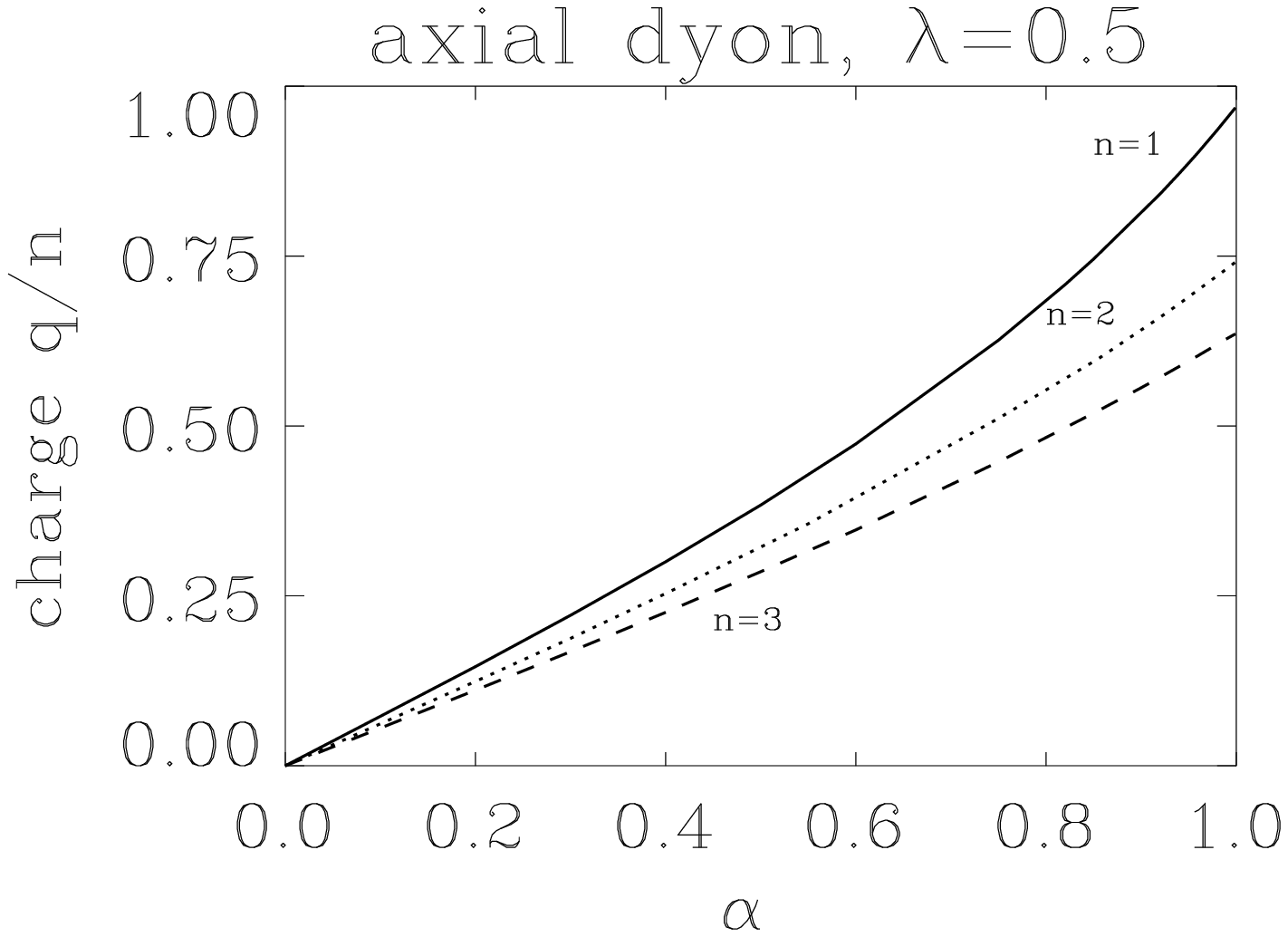}}
\caption{
The electric charge
per topological charge $q_{n}/n$
in units of $4 \pi /g$
is shown
as a function of the parameter $\alpha$
for $\lambda=0.5$ and $n=1$, 2 and 3.
}
\end{figure}

\newpage

\begin{figure}
\centering
\epsfysize=11cm
\mbox{\epsffile{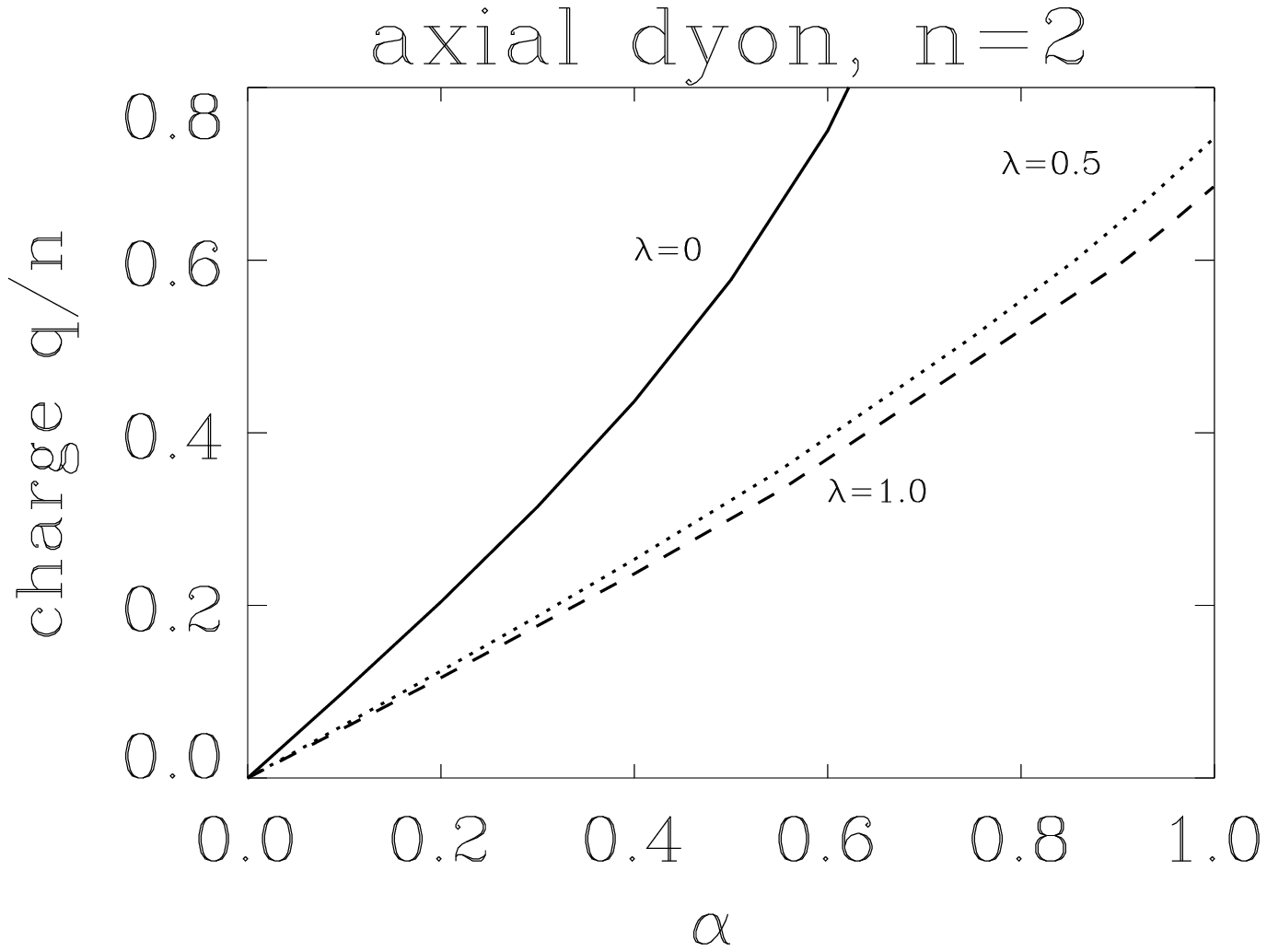}}
\caption{
The electric charge
per topological charge $q_{n}/n$
in units of $4 \pi /g$
is shown
as a function of the parameter $\alpha$
for $n=2$ for $\lambda=0$, 0.5 and 1.
}
\end{figure}

\newpage

\begin{figure}
\centering
\epsfysize=11cm
\mbox{\epsffile{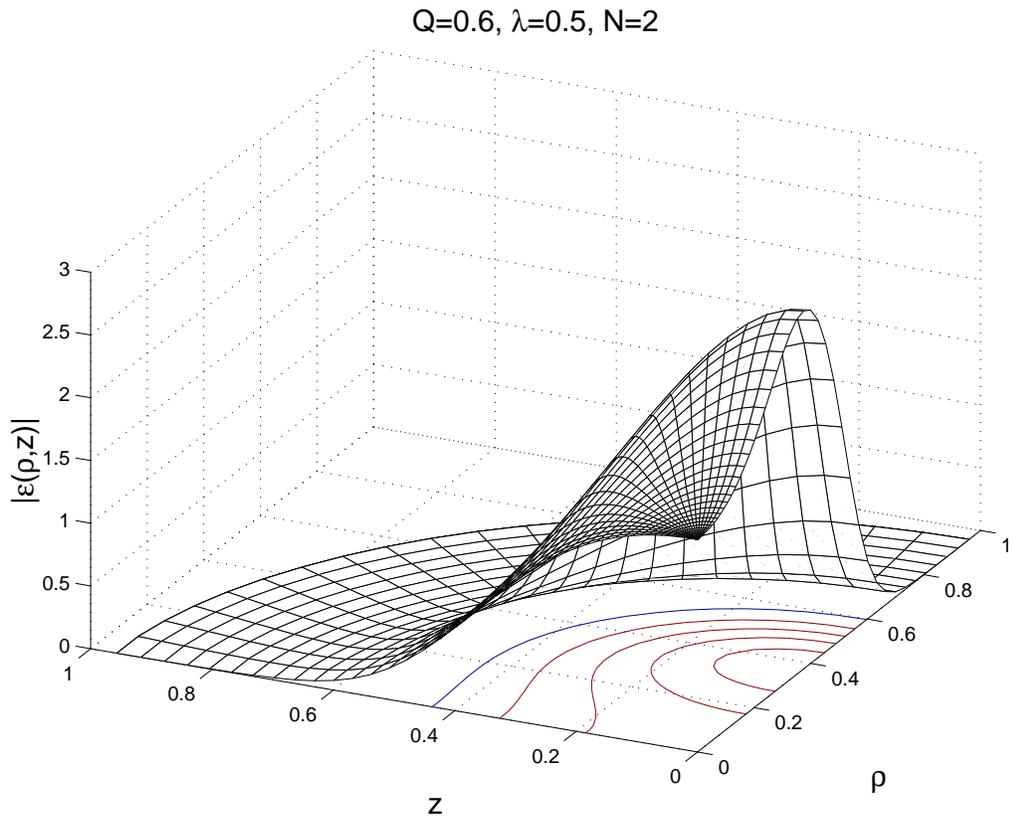}}
\caption{
The energy density $\epsilon$ 
in units of $4 \pi \eta /g$
is shown
for the dyon solution with topological charge $n=2$, electric charge
per topological charge $q_{2}/2=0.6$ and Higgs self-coupling $\lambda=0.5$
as a function of the dimensionless
compactified coordinates $\rho = \bar x \sin \theta$ and
$z = \bar x \cos \theta$, where $\bar x = x/(1+x)$.
}
\end{figure}

\newpage

\begin{figure}
\centering
\epsfysize=11cm
\mbox{\epsffile{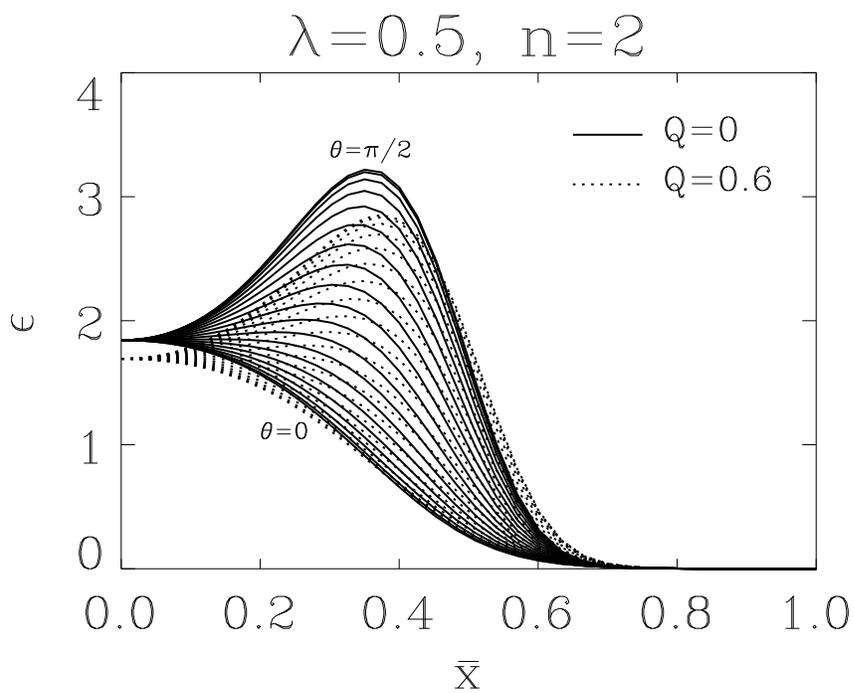}}
\caption{
The energy density $\epsilon$ 
in units of $4 \pi \eta /g$
is shown
for the dyon solution of Fig.~4 
and for the corresponding multimonopole solution
as a function of the dimensionless
compactified coordinate $\bar x$
for several values of the angle $\theta$.
}
\end{figure}

\end{document}